\documentclass[12pt]{article}
\usepackage{amsmath}
\usepackage{amssymb}
\usepackage[english]{babel}
\usepackage{latexsym}
\usepackage{times}
\usepackage{mathptm}
\begin{document}
\protect\newtheorem{principle}{Principle}[section]
\protect\newtheorem{theo}[principle]{Theorem}
\protect\newtheorem{prop}[principle]{Proposition}
\protect\newtheorem{lem}[principle]{Lemma}
\protect\newtheorem{co}[principle]{Corollary}
\protect\newtheorem{de}[principle]{Definition}
\newtheorem{ex}[principle]{Example}
\newtheorem{rema}[principle]{Remark}
\begin{titlepage} \enlargethispage{4cm}
\vspace*{-1.8cm}
\vskip.6in
\begin{center}
{\Large \textbf{Super Hilbert Spaces}
\\ \vskip.06in}
{}
\vskip.75in
{{\large  Oliver Rudolph} $^a$}
\vskip.3in
{\small \sf Starlab nv/sa}
\vskip.05in
{\small \sf Boulevard Saint Michel 47}
\vskip.05in
{\small \sf B-1040 Brussels}
\vskip.05in
{\small \sf Belgium}
\vskip.75in
\end{center}
\vskip1.56in \begin{center}
\end{center}
\vskip1in
\normalsize
\begin{center}
{ABSTRACT}
\end{center}
\smallskip
\noindent The basic mathematical framework for super Hilbert
spaces over a Gra\ss{}mann algebra with a Gra\ss{}mann number-valued
inner product is formulated.
Super Hilbert spaces over infinitely generated
Gra\ss{}mann algebras arise in the functional
Schr\"odinger representation of spinor quantum field theory
in a natural way.
\\ \bigskip \noindent
\centerline{\vrule height0.25pt depth0.25pt width4cm
\hfill}
\noindent
{\footnotesize $^a$ email: rudolph@starlab.net} \\
\end{titlepage}
\newpage
\section{Introduction}
The purpose of this article is to define and study the notion
of super Hilbert space
in a mathematically proper way and to establish generalizations of
some basic results of the theory of Hilbert spaces for super Hilbert
spaces. According to our definition a super Hilbert space is a module
over a Gra\ss{}mann algebra endowed with a Gra\ss{}mann number-valued
inner product.

The notion of super Hilbert space has first been considered by
DeWitt \cite{DeWitt92}.
Our definition, though being more general than DeWitt's
definition (see Sections 4 and 7 below), is motivated by DeWitt's work.
We shall see that DeWitt's definition is not general enough for
certain physical applications and in particular we shall see that his
notion of \emph{physical observable} on a super Hilbert space
is in general not
well-defined. Within the framework developed in this paper we
shall arrive at a more transparent notion of physical observable for
super Hilbert spaces.

In standard complex quantum theory the physical transition
amplitudes are given by the complex-valued
inner product on the underlying
complex Hilbert space.
For quantum field theories with fermionic degrees of freedom or
supersymmetric quantum theories super Hilbert spaces
Gra\ss{}mann number-valued inner products may be introduced as long as
a prescription to compute physical probabilities and transition
amplitudes is given alongside.

In the functional Schr\"odinger representation of spinor
quantum field theory super Hilbert spaces with
Gra\ss{}mann number-valued inner products arise
naturally, see \cite{Hatfield92} and below. Therefore super
Hilbert spaces provide a means to bring quantum (field) theories with
fermionic degrees of freedom and supersymmetric quantum
(field) theories into a form resembling standard quantum mechanics
which is of potential interest in several branches of quantum field
theory and may shed new light on some technical and
conceptual issues in quantum
field theory. It is the aim of the present investigation
to develop and study this notion more thoroughly.

This paper is organized as follows: in Section 2 we review some
facts about Gra\ss{}mann algebras; in Section 3 we define and
discuss the notion of a Hilbert module over a Gra\ss{}mann algebra; in
Section 4 we define super Hilbert spaces and in Section 5 we study
physical observables on super Hilbert spaces. In Section 6 we
review the Schr\"odinger representation of spinor quantum field
theory as a simple example for a situation
where a super Hilbert space arises naturally as the
state space in a quantum field theory. In Section 7 we briefly
review definitions of the notion of super Hilbert space previously
given by other authors and discuss the relation of these
definitions with our approach.

\section{Gra\ss{}mann algebras}
The Gra\ss{}mann algebra (or exterior algebra) $\Lambda_n$ with  $n$
generators is the algebra (over $\mathbb{C}$)
generated by a set of $n$ anticommuting generators $\{ \xi_i \}_{i=1}^{n}$
and by $1 \in \mathbb{C}$
\[ \xi_i \xi_j = - \xi_j \xi_i, \mbox{ for all } i,j. \]
The Gra\ss{}mann algebra generated by a
countably infinite set of generators
will be denoted by $\Lambda_\infty$. In the sequel we shall write
$\Lambda_n$ where $n \in \mathbb{N} \cup \{ \infty \}$ is possibly infinite
unless indicated otherwise.
Let $M_n := \{ (m_1, \cdots, m_k) \, \vert \, 1 \leq k \leq n,
m_i \in \mathbb{N},
1 \leq m_1 < \cdots < m_k \leq n \}$.
A special basis of $\Lambda_n$ is given by the set of elements
of the form $\xi_{m_1} \xi_{m_2} \cdots \xi_{m_k}$, with
$(m_1, \cdots, m_k) \in M_n$, together with the unit $1 \in
\mathbb{C}$.
Gra\ss{}mann algebras are
more fully discussed in, e.g., \cite{Federer69,Whitney57}.
We define an involution $^*$ on $\Lambda_n$, i.e.,
a map $^* : \Lambda_n \to \Lambda_n$ satisfying $(q^*)^* = q$, $(qp)^*
= p^* q^*$ and $(\alpha q)^* = \alpha^* q^*$
for $q,p \in \Lambda_n, \alpha
\in \mathbb{C}$ by setting $1^* := 1, \xi_i^* := \xi_i$ for all
$i$ and by extending $^*$ to all of $\Lambda_n$ by linearity.

The Gra\ss{}mann algebra carries a natural $\mathbb{Z}_2$-grading:
$\Lambda_n
= \Lambda_{n,0} \oplus \Lambda_{n,1}$, where $\Lambda_{n,0}$ consists of
the \emph{even} (commuting) elements in $\Lambda_n$ and $\Lambda_{n,1}$
consists of the \emph{odd} (anticommuting) elements of
$\Lambda_n$, i.e., for $a_r \in \Lambda_{n,r}$ and $a_s \in \Lambda_{n,s}$
we have $a_r a_s = (-1)^{rs}a_s a_r \in \Lambda_{n, r+s(\mathrm{mod} 2)}$.
We also write $\deg(a_r) = r$ if $a_r \in \Lambda_{n, r}$ and call
$\deg(a_r)$ the \emph{degree} of $a_r$.
Every element $q \in \Lambda_n$ can be uniquely decomposed as
\begin{equation} \label{body}
q = q_B 1 + q_S =
q_B 1 + \sum_{(m_1, \cdots, m_k) \in M_n} q_{m_1, \cdots,
m_k} \xi_{m_1} \cdots \xi_{m_k}, \end{equation}
where $q_B, q_{m_1, \cdots, m_k} \in \mathbb{C}$.
The complex number $q_B$ is
called the \emph{body} of $q$ and the Gra\ss{}mann number
$q_S$ is called the \emph{soul} of $q$. Notice that the body
operation is linear and respects the algebra structure, i.e., $(q+p)_B =
q_B + p_B$ and $(pq)_B = p_B q_B$ for all $q,p \in \Lambda_n$.

The Gra\ss{}mann algebra can also be written as a direct sum
\[ \Lambda_n = \oplus_{r=0}^n \mathsf{V}_r, \]
where $\mathsf{V}_r$ is the complex
vector space spanned by the elements of the
form $\xi_{m_1} \cdots \xi_{m_r}$, $r$ fixed.
Therefore any $q \in \Lambda_n$
can be uniquely decomposed as $q = \sum_{r=0}^n q_r$ with $q_r \in
\mathsf{V}_r$. Any choice of a basis of
$\mathsf{V}_1$ may serve as a possible choice of (possibly complex)
generators of $\Lambda_n$.

For $n$ finite there is an isomorphism $\star : \Lambda_n \to \Lambda_n$
known as the \emph{Hodge star operator}. Consider the ordered sequence
$\{ \xi_1, \cdots, \xi_n \}$ of all generators of $\Lambda_n$, then $\star$
is defined on the element $\xi_{i_1} \xi_{i_2} \cdots \xi_{i_d}$ by
$\star[\xi_{i_1} \xi_{i_2} \cdots \xi_{i_d}] := \xi_{j_1} \xi_{j_2} \cdots
\xi_{j_{n-d}}$, where $(j_1, \cdots, j_{n-d})$ is chosen such that
$(i_1, \cdots, i_d, j_1, \cdots, j_{n-d})$ is an even permutation
of $(1, \cdots, n)$. We extend $\star$ to all of $\Lambda_n$ by
conjugate linearity, i.e., we require that $\star[\alpha q] := \alpha^*
\star[q]$ for $\alpha \in \mathbb{C}$ and $q \in \Lambda_n$ and that
$\star$ is a real linear transformation. It is well-known
that the Hodge star operator is independent of the basis used to
define it.

Now expand every $q \in \Lambda_n$ with respect to the basis
of $\Lambda_n$ as in Equation \ref{body}.
Then we can define for each $1 \leq \kappa < \infty$
\begin{equation}
\vert q \vert_\kappa := \left( \vert q_B \vert^\kappa + \sum_{(m_1,
\cdots, m_k) \in M_n} \vert q_{m_1, \cdots, m_k} \vert^\kappa
\right)^{1/\kappa}. \label{norm} \end{equation}
Moreover, we define $\vert q \vert_\infty
:= \sup_{(m_1, \cdots, m_k) \in M_n} \vert q_{m_1, \cdots, m_k}
\vert$. If $n$ is finite, it is straightforward to
verify that each $\vert \cdot
\vert_\kappa$ defines a norm on $\Lambda_n$ and that
$\Lambda_n$ becomes a
complex Banach space with each of the norms $\vert \cdot \vert_\kappa$,
$1 \leq \kappa
\leq \infty$, which we denote by $\Lambda_n(\kappa)$ respectively.
In the case of $\Lambda_\infty$, $\vert \cdot \vert_\kappa$
defines a seminorm on $\Lambda_\infty$ and
 we denote the set of all $q \in
\Lambda_\infty$ for which the above expression for $\vert q \vert_\kappa$
satisfies $\vert q \vert_\kappa < \infty$ by $\Lambda_\infty(\kappa)$.
Again
it is easy to see that $\Lambda_\infty(\kappa)$ with the norm $\vert \cdot
\vert_\kappa$ is a Banach space for all $1 \leq \kappa \leq \infty$.
The norm $\vert \cdot \vert_1$ is sometimes also called the
\emph{Rogers norm} and $\Lambda_n(1)$ the \emph{Rogers algebra},
see \cite{Rogers80}.

The norms $\vert \cdot \vert_\kappa$ in (\ref{norm})
depend implicitly on the
choice of the set of generators of the Gra\ss{}mann algebra and are
not invariant under a change of the set of generators of $\Lambda_n$.
Gra\ss{}mann number-valued variables appearing in physical  theories
are by their very nature
not directly observable and therefore in general there may be no
physically preferred choice for the set of generators of a Gra\ss{}mann
algebra.
However, for $n$ finite, it is well-known that
not only all the norms in (\ref{norm})
are equivalent and therefore generate the same topology on
$\Lambda_n$ for all $1 \leq \kappa \leq \infty$ but the resulting
topology is in fact independent of the choice of generators of the
Gra\ss{}mann algebra (this is an immediate consequence of
Proposition 1.2.16 in \cite{KadisonR86}).
It is evident that the Hodge star operator
is continuous in this topology.

There is a norm, invariant under change of generators,
on $\Lambda_n$ which can be constructed as follows.
Firstly, it is known that there is a norm $\Vert \cdot \Vert_r$
on $\mathsf{V}_r$ given by, \cite{Federer69,Whitney57},
\begin{equation}
\Vert q_r \Vert_r = \inf \left\{ \sum_{(m_1,
\cdots, m_r) \in M_n} \vert q_{m_1, \cdots, m_r} \vert
\right\}, \end{equation} for $q_r \in \mathsf{V}_r$
where the infimum is taken over all possible choices of the set of
generators of the Gra\ss{}mann algebra.
The norm $\Vert \cdot \Vert_r$ satisfies $\Vert q_r p_s \Vert_{r+s}
\leq \Vert q_r \Vert_r \Vert p_s \Vert_s,$ for all $q_r \in
\mathsf{V}_r$ and $p_s \in \mathsf{V}_s$, see \cite{Federer69,Whitney57}.
Now define a seminorm on $\Lambda_n$ by
\begin{equation}
\Vert q \Vert := \sum_{r=0}^n \Vert q_r \Vert_r. \label{norm2}
\end{equation}
For $n$ finite it is obvious that $\Vert \cdot \Vert$ is a norm on
$\Lambda_n$. This norm $\Vert \cdot \Vert$ is called the
\textbf{mass (norm)} on $\Lambda_n$ ($n$ finite). By construction
the mass norm is
independent of the choice of the set of generators of $\Lambda_n$.

If $n = \infty$, then every finite subset $\{ \xi_{i_1}, \cdots
\xi_{i_m} \} \cup \{ 1 \}$ of the set of all generators
$\{ \xi_i \}_i$ of
$\Lambda_\infty$ generates an $m$-dimensional Gra\ss{}mann
subalgebra of $\Lambda_\infty$
denoted by $\Lambda_{i_1, \cdots, i_m}$. The collection of
all such Gra\ss{}mann subalgebras of $\Lambda_\infty$
forms a directed set and the canonical
imbedding morphisms obviously preserve the mass norm.
We consider the algebraic direct limit $\Delta_\infty$ of this
directed set. The mass norm on the finite dimensional
Gra\ss{}mann subalgebras induces a \emph{mass norm} $\Vert \cdot \Vert$
on $\Delta_\infty$.
We denote the completion of $\Delta_\infty$ with respect to the mass
norm by $\Lambda_\infty^m$. Obviously, $\Lambda_\infty^m$ consists
of all $q \in \Lambda_\infty$
with $\Vert q \Vert = \sum_{r=0}^\infty \Vert q_r \Vert_r < \infty$.
The norm on $\Lambda_\infty^m$ is again called the \textbf{mass norm}.

It is easy to see that the mass
norm is submultiplicative $\Vert pq \Vert =  \sum_r \Vert (pq)_r \Vert_r
\leq \sum_r \sum_{k \leq r} \Vert p_{r-k} q_k \Vert_r \leq
\sum_r \sum_{k \leq r} \Vert
p_{r-k} \Vert_{r-k} \Vert q_k \Vert_k \leq \sum_r \sum_k \Vert p_r
\Vert_r \Vert q_k \Vert_k = \Vert p \Vert \Vert q \Vert$.

Notice that there is a seminorm on $\Lambda_n$ which is
trivially independent of the choice of the set of generators,
namely \begin{equation}
\Vert q \Vert_B := \vert q_B \vert.
\end{equation}

We have $\Vert q \Vert_B \leq \vert q \vert_\kappa$ for all
$q \in \Lambda_n(\kappa)$ and $\Vert q \Vert_B \leq \Vert q \Vert$
for all $q \in \Lambda_\infty^m$. \\

In the sequel it is understood that the symbol $\Lambda$ stands
for either (a) $\Lambda_n$ with $n$ finite, (b) for $\Lambda^m_{\infty}$,
or (c) for $\Lambda_\infty(\kappa)$ with $1 \leq \kappa \leq
\infty$.

\section{Hilbert $\Lambda$ modules}
\begin{de}
A \textbf{pre-Hilbert} $\Lambda$ \textbf{module} is a
$\mathbb{Z}_2$-graded right $\Lambda$
module $E = E_0 \oplus E_1$ equipped with
a $\Lambda$-valued inner product
$\langle \cdot, \cdot \rangle : E \times E \to \Lambda$ that is
sesquilinear, definite, and whose body is Hermitean and
positive. In other words:
\begin{enumerate}
\item $\langle x, y_1 + y_2 \rangle = \langle x, y_1 \rangle + \langle x,
y_2 \rangle$, and $\langle y_1 + y_2, x \rangle = \langle y_1, x \rangle
+ \langle y_2, x \rangle$ for $x,y_1,y_2 \in E$;
\item $\langle x, \alpha y \rangle = \alpha \langle x, y \rangle =
\langle \alpha^* x, y \rangle$,
for $x,y \in E, \alpha \in \mathbb{C}$;
\item $\langle x, y \rangle_B = \langle y, x \rangle_B^*$, for $x,y \in
E$;
\item $\langle x, x \rangle_B \geq 0$ for $x \in E$ and $\langle x, x
\rangle =0$ if and only if $x = 0$. \end{enumerate} \label{def2}
\end{de}
An element $x$ of a pre-Hilbert $\Lambda$ module $E = E_0 \oplus E_1$
is called \emph{even} if $x \in E_0$ and \emph{odd} if $x \in E_1$,
respectively.

Immediate consequences of Definition \ref{def2} are that every
pre-Hilbert $\Lambda$ module is a complex vector space and that
every element $x$ of a pre-Hilbert $\Lambda$ module $E$ can be
uniquely written as a sum of an even and an odd element of $E$,
i.e., $x= x_0 + x_1$, where $x_0 \in E_0$ and $x_1 \in E_1$.

\begin{rema} \label{3.2}
The rationale behind the positivity requirement 4 in Definition \ref{def2}
is to interpret the body of  the inner product
$\langle \cdot, \cdot \rangle$ as physical transition amplitude.
\end{rema} \begin{rema}
DeWitt \cite{DeWitt92}
requires the inner product on a super Hilbert space to be
sesquilinear with respect to Gra\ss{}mann numbers, i.e.,
$\langle x, y \rangle q = \langle x, y q \rangle$, for all $x, y$ in the
super Hilbert space and $q \in \Lambda_n$. We shall see below, however,
that the inner product on the super Hilbert space arising in the
functional Schr\"odinger representation of
spinor quantum field theory does not satisfy this condition.
Accordingly we have allowed for a more general notion of pre-Hilbert
$\Lambda$ module. \end{rema}

We may now use a norm $\Vert \cdot \Vert_\Lambda$
defined on $\Lambda$ to define a norm $\Vert \cdot \Vert_E$
on a pre-Hilbert $\Lambda$ module $E$ by
\begin{equation} \Vert x \Vert_E^2 =
\Vert \langle x, x \rangle \Vert_\Lambda. \label{knorm}
\end{equation}
For instance, if $\Lambda$ equals $\Lambda_n$ or
$\Lambda_\infty(\kappa)$ endowed with the norm $\vert \cdot
\vert_\kappa$, then the norm on $E$ is given by
\begin{equation}
\Vert x \Vert_\kappa^2 := \vert \langle x, x \rangle
\vert_\kappa, \end{equation} for $x \in E$ and $1 \leq \kappa \leq \infty$.
The norm \begin{equation}
\Vert x \Vert^2 := \Vert \langle x, x \rangle \Vert \end{equation}
corresponding to the mass norm on $\Lambda = \Lambda_n$ or
$ \Lambda = \Lambda_\infty^m$
in Equation \ref{norm2} is called the \textbf{mass
norm} on the Hilbert $\Lambda$ module $E$.

In the sequel it is understood that we consider only
norms on a Hilbert $\Lambda$ module arising from a norm on $\Lambda$
as in Equation \ref{knorm} unless explicitly stated otherwise.
\begin{lem}[Cauchy-Schwarz inequality]
\label{lem1}
If $E$ is a pre-Hilbert $\Lambda$ module and $x,y \in
E$, then \[ \vert \langle x, y \rangle_B \vert^2 \leq
\langle x, x \rangle_B \langle y, y \rangle_B. \]
\end{lem}
\emph{Proof}: Let $p_x := \langle x, x \rangle_B, p_y := \langle y, y
\rangle_B, q := \langle x, y \rangle_B$ and $\lambda \in \mathbb{R}$, then
\[ 0 \leq \langle x - y \lambda q^*, x - y \lambda q^* \rangle_B =
p_x - 2 \lambda q q^* + \lambda^2 q p_y q^*. \]
Adding $2 \lambda q q^*$ on both sides and taking norms yield
\begin{equation} \label{eq1}
2 \lambda \vert q \vert^2
\leq 2 \vert \lambda \vert \vert q \vert^2 \leq \vert p_x
+ \lambda^2 q p_y q \vert
\leq \vert p_x \vert + \lambda^2 \vert q
\vert^2 \vert p_y \vert. \end{equation}
This is equivalent to \[ \left( \lambda \vert q \vert \vert
p_y \vert - \vert q \vert \right)^2 \geq (\vert q
\vert)^2 - \vert p_x \vert \vert p_y \vert. \]
If $\vert p_y \vert \neq 0$, then
setting $\lambda := \frac{1}{\vert p_y \vert}$ yields the required
inequality. Moreover, we find that $\vert p_x \vert = 0$ and $\vert p_y
\vert \neq 0$ implies $\vert q \vert = 0$ (let $\lambda = 1)$.
From symmetry considerations (or from Equation \ref{eq1})
we also get that $\vert p_y \vert = 0$
and $\vert p_x \vert \neq 0$ implies $\vert q \vert =0$.
In the case that $\vert p_x \vert = \vert p_y \vert =0$ we infer
from Equation \ref{eq1} by taking $\lambda$ to be positive that $\vert q
\vert = 0$.  $\Box$ \\

On any pre-Hilbert $\Lambda$ module $E$ there is a \emph{body
operation}, i.e., a linear map $B : E \to E_0, x \mapsto x_B$ such
that $(x \lambda)_B = x_B \lambda_B$ for all $\lambda \in \Lambda$
\cite{NagamachiK92}. First define the \emph{soul} $s(E)$ and the \emph{body}
$b(E)$ of $E$ by \begin{eqnarray*}
s(E) & := & \{ x \in E \vert x \lambda = 0 \mbox{ for
some } \lambda \in \Lambda, \lambda \neq 0 \}, \\
b(E) & := &  E/s(E). \end{eqnarray*}
The body operation $B : E \to E_0$ is the canonical
surjection from $E$ to $b(E)$.

If the inner product of $E$ satisfies $\langle
x_B, y_B \rangle = \langle x, y \rangle_B$, then the body of $E$
endowed with the induced inner product is a pre-Hilbert space
whose completion is a Hilbert space (by virtue of the Cauchy-Schwarz
inequality). But even if
the inner product does not respect the body operation, we can prove
\begin{prop} \label{3.3}
Let $E$ be a pre-Hilbert $\Lambda$ module. Then there exists a map $x \to
[x]$ from $E$ into a dense subspace of a Hilbert space $H$ such that
\[ \langle [x], [y] \rangle_H = \langle x, y \rangle_B, \] for all $x,y \in
E,$ where $\langle \cdot, \cdot \rangle_H$ denotes the inner product
on $H$. \end{prop}
\emph{Proof}: Let $\mathcal{N} := \{ x \in E \vert \langle x, x
\rangle_B
= 0 \}$. Let $[x] := x + \mathcal{N}$. Then $\langle \cdot, \cdot
\rangle_B$ induces a well-defined inner product on $E/\mathcal{N}$
by virtue of Lemma \ref{lem1}. Therefore $E/\mathcal{N}$ with this
inner product is a pre-Hilbert space.  $\Box$
\begin{de}
Let $E$ be a pre-Hilbert $\Lambda$ module and $\Vert \cdot \Vert$
a norm on $E$, then $E$ is said to be a \textbf{Hilbert} $\Lambda$
\textbf{module} if $E$ is complete with respect to its norm. A
\textbf{Hilbert submodule} of a Hilbert module $E$ is a closed
submodule of $E$. \end{de} \begin{de}
Let $E$ and $F$ be Hilbert $\Lambda$ modules.
A $\mathbb{C}$-linear map $O : E \to E$ is called an
\textbf{operator} on $E$.
We denote the set of all bounded operators on $E$ by $\mathcal{L}(E)$.
An operator $T : E \to E$ is called \textbf{unitary} if $\langle
T(x), T(y) \rangle = \langle x, y \rangle$ for all $x,y \in
E$. An operator $S$ is called \textbf{weakly unitary} if $\langle
S(x), S(y) \rangle_B = \langle x, y \rangle_B$ for all $x,y \in
E$. \label{def3}
A \textbf{(Hilbert)
module map} is a linear map $T : E \to F$ which respects the
module action: $T(x q) = T(x) q$, for $x \in E, q \in
\Lambda$. \end{de} \begin{de}
A Hilbert $\Lambda$ module $E$ is said to satisfy the \textbf{strong
definiteness condition} if $\langle x, x \rangle_B = 0$ implies $x
=0$ for all $x \in E$. \end{de}
Every Hilbert $\Lambda$ module $E$ satisfying the strong
definiteness condition becomes a pre-Hilbert space with respect to
the norm $\Vert \cdot \Vert^2_B := \langle \cdot, \cdot \rangle_B.$
\\

Every Hilbert $\Lambda$ module $E$ is endowed with a
$\mathbb{Z}_2$-grading $E = E_0 \oplus E_1$. This induces a
$\mathbb{Z}_2$-grading on $\mathcal{L}(E)$: every operator
$T : E \to E$ can be written as sum of
an \emph{even} map $T_0 : E_i \to E_i$ and an \emph{odd} map $T_1 :
E_i \to E_{i+1 (\mathrm{mod} 2)}$, i.e. $T= T_0 + T_1$ where $T_0$
and $T_1$ are
defined by $T_0u := (Tu_0)_0 + (Tu_1)_1$ and $T_1u := (Tu_0)_1 +
(Tu_1)_0$ respectively where $u = u_0 + u_1$.
\begin{de} \label{def4}
Let $E$ be a Hilbert $\Lambda$ module. An operator $T : E \to E$ is
said to be \textbf{adjointable} if there exists an operator $T^* : E \to
E$ satisfying $\langle x, T y \rangle = \langle T^* x, y \rangle$
for all $x,y \in E$. Such an operator $T^*$ is called an
\textbf{adjoint} of $T$. We denote the set of all adjointable operators
on $E$ by $\mathfrak{B}(E)$. An adjointable operator $T \in
\mathfrak{B}(E)$ is called \textbf{self-adjoint} if $T^* = T$. \\
An operator $T : E \to E$ is
said to be \textbf{weakly adjointable} if there exists an operator
$T^\dagger : E \to
E$ satisfying $\langle x, T y \rangle_B = \langle T^\dagger x, y \rangle_B$
for all $x,y \in E$. Such an operator $T^\dagger$ is called a
\textbf{weak adjoint} of $T$. We denote the set of all weakly
adjointable operators
on $E$ by $\mathfrak{B}_w(E)$. A weakly adjointable operator $T \in
\mathfrak{B}_w(E)$ is called \textbf{weakly self-adjoint} if $T^\dagger
= T$. \end{de} \begin{rema}
Obviously, any adjointable operator is also weakly adjointable. Thus,
$\mathfrak{B}(E)\subset \mathfrak{B}_w(E)$. We have noticed above
in Remark \ref{3.2} that the body of the inner product on a
Hilbert $\Lambda$ module is interpreted as the physical transition
amplitude. Accordingly we also expect that the set
$\mathfrak{B}_w(E)$ plays a distinguished role and that the operators
representing physical observables or physical operations will be
elements of $\mathfrak{B}_w(E)$.
\end{rema}
The following proposition can be proven in analogy to the
corresponding result for Hilbert $C^*$-modules, see
\cite{WeggeOlsen93}.
\begin{lem} (a)
Let $E$ be a Hilbert $\Lambda$ module and $T : E \to E$
be an adjointable operator. The adjoint $T^*$ of $T$ is unique.
If both $T: E \to E$ and $S: E \to E$ are adjointable operators, then $ST$
is adjointable and $(ST)^* = T^* S^*$. \\
(b) Let $E$ be a Hilbert $\Lambda$ module
satisfying the strong definiteness
condition and $T_w : E \to E$
be a weakly adjointable operator. Then the weak adjoint $T_w^\dagger$ of
$T_w$ is unique.
If both $T_w: E \to E$ and $S_w: E \to E$ are adjointable operators, then
$S_wT_w$ is adjointable and
$(S_wT_w)^\dagger = T^\dagger_w S^\dagger_w$.
\end{lem}
\emph{Proof}: (a) Assume that $\overline{T}$ and $T^*$ are adjoints
of $T$, then \[ 0 = \langle \overline{T} x, y \rangle - \langle
T^* x, y \rangle = \langle (\overline{T} - T^* ) x, y \rangle, \]
for all $x,y \in E$. Let $y = (\overline{T} - T^*) x$. This
implies $\overline{T} = T^*$. A similar argument proves (b).
 $\Box$

\section{Super Hilbert spaces}
The Definitions \ref{def3} and \ref{def4}
are analogous to parallel definitions in the theory of Hilbert
$C^*$-modules \cite{WeggeOlsen93,Lance95}. However,
the positivity requirement in the definition of a Hilbert $\Lambda$
module is weaker than the positivity requirement for Hilbert
$C^*$-modules and all results for Hilbert $C^*$-modules depending on the
positivity of the inner product may in general not be valid for a
Hilbert $\Lambda$ module. The Cauchy-Schwarz inequality in Lemma
\ref{lem1} is a first example. As a consequence of the failure of the
general Cauchy-Schwarz inequality the inner product
on a pre-Hilbert $\Lambda$ module may in general not be
continuous in each argument and therefore in general
an inner product on a
pre-Hilbert $\Lambda$ module does not extend to an inner product
on its completion. In the sequel we shall be mainly interested in
inner products which are continuous.
\begin{de} \label{SH}
We shall call a (pre-) Hilbert $\Lambda$ module $\mathcal{H}$
a \textbf{super (pre-) Hilbert space} if the inner product on $\mathcal{H}$
is continuous, i.e., if there exists a
constant $C>0$ such that
$\Vert \langle x, y \rangle \Vert \leq C \Vert x \Vert \Vert y \Vert$.
\end{de} \begin{rema}
The completion of a super pre-Hilbert space is a super Hilbert
space. \end{rema}
\begin{rema}
All concrete examples for super Hilbert spaces we shall discuss below will
satisfy the strong definiteness condition. An example for a
situation where a super Hilbert space without the strong
definiteness condition arises is the Becchi-Rouet-Stora-Tyutin (BRST)
formulation of gauge theories.
The natural choice of the state space arising the BRST
formulation of gauge theories is a Hilbert $\Lambda$ module
(as there is always a representation of the Gra\ss{}mann algebra acting
on the state space) endowed with an indefinite inner product.
The \emph{physical} states are annihilated by the BRST operator
$\Omega$, i.e., satisfy the condition $\Omega \psi_{phys} =
0$. The inner product induced on the set $\mathcal{V}_{phys}$ of
all physical states can be shown to be
positive but not definite. The states in $\mathcal{V}_{phys}$
with probability zero are called \emph{ghost states} and are
unobservable. Therefore in the BRST formulation of gauge theories
we naturally arrive at a physical state space which  carries the
structure of Hilbert $\Lambda$ module or a super Hilbert space not
satisfying the strong definiteness condition. A good
introduction into the BRST formalism can be found, e.g., in
\cite{Kugo97}. \end{rema}
We  have already noticed above that the physical transition amplitudes
are given by the body of the inner product
of a Hilbert $\Lambda$ module. This gives rise to the following
definition. \begin{de}
Let $\mathcal{H}$ be a super Hilbert space. An element $x \in \mathcal{H}$
is called \textbf{physical} if $\langle x,  x \rangle_B \neq 0.$
An element $g \in \mathcal{H}$ with $g \neq 0$ and $\langle g, g \rangle_B
= 0$ is called a \textbf{ghost}.
\end{de} \begin{ex} \label{ex1}
Let $n$ be finite. The Gra\ss{}mann algebra $\Lambda_n$ endowed
with the mass norm $\Vert \cdot \Vert$ becomes a super Hilbert
space with the inner product $\langle \cdot, \cdot \rangle$
given by \begin{equation}
\label{Eq2} \langle p, q \rangle := \star \left[ p \star[q] \right]
\end{equation}
for all $p,q \in \Lambda_n$, where $\star$ denotes the Hodge star operator.
The submultiplicativity of the mass norm implies
$\Vert \langle p, q \rangle \Vert \leq \Vert p \Vert
\Vert q \Vert$ for all $p,q \in \Lambda_n$.
Recalling Equation \ref{body} \[ q =
q_B 1 + \sum_{(m_1, \cdots, m_k) \in M_n} q_{m_1, \cdots,
m_k} \xi_{m_1} \cdots \xi_{m_k}, \] we see that \[ \langle q, q
\rangle_B = \vert q_B \vert^2 + \sum_{(m_1, \cdots, m_k) \in M_n}
\vert q_{m_1, \cdots, m_k} \vert^2. \] Therefore $\Lambda_n$ with
the inner product (\ref{Eq2})
satisfies the strong definiteness condition.

More general super Hilbert spaces can be constructed by building
the tensor product $\Lambda_n \otimes \mathfrak{H}$
of $\Lambda_n$ with a complex Hilbert space $\mathfrak{H}$. The
inner product of $\Lambda_n \otimes \mathfrak{H}$ is given on
simple tensors by $\langle p \otimes \varphi, q \otimes \psi \rangle =
\langle p, q \rangle \langle \varphi, \psi \rangle$, for $p, q \in
\Lambda_n$ and $\varphi, \psi \in \mathfrak{H},$
and extended
to arbitrary elements of $\Lambda_n \otimes \mathfrak{H}$ by
linearity and continuity. We omit the details of the construction as a
more general example will be given below in Example \ref{4.4}.
\end{ex}
\begin{ex} \label{exa}
Consider a measure space $(X, \Omega)$, where $X$ is a set and $\Omega$
a $\sigma$-algebra of subsets of $X$, endowed with a $\sigma$-finite
measure $\mu$. Every function $f : X \to \Lambda_n$ can be expanded
as \[ f(x) = f_B(x) + \sum_{(m_1, \cdots, m_k) \in M_n} f_{m_1, \cdots,
m_k}(x) \xi_{m_1} \cdots \xi_{m_k}, \] with complex-valued
functions $f_B : X \to \mathbb{C}$ and
$f_{m_1, \cdots, m_k} : X \to \mathbb{C}$. We restrict ourselves here
to the case that $n$ is finite. Now consider the set $E$ of all
functions $f : X \to \Lambda_n$ such that $f_B$ and all $f_{m_1, \cdots,
m_k}$ are square integrable with
respect to $\mu$. This requirement is independent of
the basis chosen. We define a
$\Lambda_n$-valued inner product on $E$
by \begin{equation} \label{kinner}
\langle f, g \rangle = \int f(x)^* g(x) d \mu(x),
\end{equation} for all $f, g \in E$.
If $\Lambda_n$ is furnished with the Rogers norm $\vert \cdot \vert_1$,
then define \begin{equation} \label{knorm2}
\Vert f \Vert :=
\sum_{(m_1, \cdots, m_r) \in M^0_n} \sqrt{ \int \vert f_{m_1, \cdots, m_r}(x)
\vert^2 d \mu(x)},
\end{equation} where we introduced the notation
$M_n^0 := \{ (m_1, \cdots, m_k) \, \vert \, 0 \leq k \leq n,
m_i \in \mathbb{N},
1 \leq m_1 < \cdots < m_k \leq n \}$ and $f_\emptyset :=
f_B$. Further let $\mathcal{N} := \{ f \in E \, \vert \, \Vert f \Vert
=0 \}$. It is easy to see that Equation \ref{knorm2} defines
a norm on $E/\mathcal{N}$ and that $E/\mathcal{N}$ equipped with the norm
(\ref{knorm2}) becomes a super Hilbert space. Indeed, let $f, g \in
E$, then \begin{eqnarray*}
\vert \langle f, g \rangle \vert_1 & = &
\sum_{(m_1, \cdots, m_r) \in
M_n^0} \left\vert \sum_{k=0}^r \sum_{\sigma} \!^\prime
(-1)^{\mathrm{sgn}(\sigma)}
\int f^*_{\sigma(m_1), \cdots, \sigma(m_k)} g_{\sigma(m_{k+1}),
\cdots, \sigma(m_r)} d \mu \right\vert \\
& \leq & \sum_{(m_1, \cdots, m_r) \in
M_n^0} \sum_{k=0}^r \sum_{\sigma} \!^\prime
\int \vert f_{\sigma(m_1), \cdots, \sigma(m_k)} \vert \, \vert
g_{\sigma(m_{k+1}), \cdots, \sigma(m_r)} \vert d \mu \\
& \leq & \sum_{(m_1, \cdots, m_r) \in
M_n^0} \sum_{k=0}^r \sum_{\sigma} \!^\prime
\left[{\int \vert f_{\sigma(m_1), \cdots, \sigma(m_k)} \vert^2 d
\mu
\int \vert g_{\sigma(m_{k+1}), \cdots, \sigma(m_r)} \vert^2 d
\mu} \right]^{\frac{1}{2}} \\ & \leq & \Vert f \Vert \, \Vert g \Vert,
\end{eqnarray*} where the sum $\sum'_\sigma$ in the first three lines
runs over all permutations $\sigma$
of $(m_1, \cdots, m_r)$ such that $(\sigma(m_1), \cdots,
\sigma(m_k)) \in M_n^0$ and $(\sigma(m_{k+1}),
\cdots, \sigma(m_{r})) \in M_n^0$. If we replace (\ref{kinner}) by
$\langle f, g \rangle = \int \star[f(x)] g(x) d \mu(x)$, a similar
argument holds. \end{ex}
\begin{ex}
For $n$ infinite we also can make $\Lambda_\infty^m$
a super Hilbert space by defining an
appropriate inner product.
For simplicity we assume that the set of all generators is
countable $\{ \xi_i \}_{i \in \mathbb{N}}$.
The generalization of the following to the situation
where the set of generators is uncountable is obvious.
First of all we observe that the inner
product (\ref{Eq2}) is not well-defined as the Hodge star operator
is not defined on $\Lambda_\infty^m$.
This difficulty can be overcome by suitably imbedding
$\Lambda_\infty^m$ into the direct sum $\Lambda_\infty^m \oplus
\Lambda_\infty^m$ of two copies of
$\Lambda_\infty^m$.
The basic idea is to introduce the formal infinite product of all
generators $\xi_\infty \equiv \prod_i \xi_i$. We do not make any
attempt to give a precise meaning to this infinite product of Gra\ss{}mann
numbers and just introduce $\xi_\infty$ as an auxiliary object which has
certain properties we would expect from the product of all
generators of the Gra\ss{}mann algebra. Namely, we require that $q
\xi_\infty = q_B \xi_\infty$ for all $q \in \Lambda_\infty^m$.
Analogously we define cofinite products of the generators of the
Gra\ss{}mann algebra, i.e., infinite products obtained from
$\xi_\infty$ by removing at most finitely many terms in the
product. E.g., the infinite product $\prod_{i \neq 1} \xi_i$ of
all generators except $\xi_1$ is denoted by $\hat{\xi}_1 \equiv
\frac{\partial}{\partial \xi_1} \xi_\infty.$ We require \[
\frac{\partial}{\partial \xi_i} \frac{\partial}{\partial \xi_j} = -
\frac{\partial}{\partial \xi_j} \frac{\partial}{\partial \xi_i} \]
and $\xi_i \hat{\xi}_i = \xi_\infty$ and $\xi_i
\frac{\partial}{\partial \xi_j} = - \frac{\partial}{\partial \xi_j}
\xi_i,$ for all $i \neq j$. Moreover we require $\xi_\infty$ to
be even. Therefore the algebra $\star \left[
\Lambda^m_\infty \right]$ generated by the $\frac{\partial}{\partial
\xi_i}$ and $1$ is isomorphic to $\Lambda_\infty^m$.

Now we are able to define the action of the \emph{Hodge star
operator} on $\Lambda_\infty^m$ by setting
\begin{equation} \label{hodgestar}
\star[q] \equiv q_B^* \xi_\infty + \sum_{(m_1, \cdots, m_k) \in
M_\infty} q_{m_1, \cdots, m_k}^* \frac{\partial}{\partial \xi_{m_k}} \cdots
\frac{\partial}{\partial \xi_{m_1}}
\xi_{\infty}, \end{equation} for all $q \in \Lambda_\infty^m$.
Moreover, we require $\star[\star[q]] = q,$ for all $q$.
The algebra generated by the $\frac{\partial}{\partial \xi_i}$ is
isomorphic to $\Lambda_\infty^m$ with the isomorphism given by
the Hodge star operator (\ref{hodgestar}).

The inner product $\langle p, q \rangle = \star[ p \star[q]],$
for all $p, q \in \Lambda_\infty^m$ is now well-defined. Notice
that although $\star[q] \notin \Lambda_\infty^m$ for all $q \in
\Lambda_\infty^m$, the inner product satisfies
$\langle p, q \rangle \in \Lambda_\infty^m$ if $p,q \in
\Lambda_\infty^m$. Since, by virtue of the properties of the mass
norm, we also have $\Vert \langle p, q \rangle \Vert \leq \Vert p \Vert
\Vert q \Vert$ for all $p, q \in \Lambda_\infty^m$ and since
\[ \langle q, q \rangle_B = \vert q_B \vert^2 +
\sum_{(m_1, \cdots, m_r) \in
M_n} \vert q_{m_1, \cdots, m_r} \vert^2 \]
we see that $\Lambda_\infty^m$ with the inner product (\ref{Eq2})
is a super Hilbert space satisfying the strong definiteness
condition. \end{ex}
\begin{ex}
$\star \left[ \Lambda^m_\infty \right]$ can be made a super Hilbert
space (over $\Lambda_\infty^m$)
by setting \[ \langle p, q \rangle = \star[p] q, \] for all $p, q \in
\star[\Lambda^m_\infty]$ (when we identify $\xi_\infty$ formally
with $1 \in \mathbb{C}$). Obviously $\star \left[ \Lambda_\infty^m
\right]$ satisfies the strong definiteness condition. \end{ex}
\begin{ex} We are now going to construct the tensor product of two
super Hilbert spaces $\mathcal{H}_1$ and $\mathcal{H}_2$. We
denote the inner products on $\mathcal{H}_1$ and $\mathcal{H}_2$
by $\langle \cdot, \cdot \rangle_1$ and $\langle \cdot, \cdot
\rangle_2$ respectively, and the norms on $\mathcal{H}_1$ and
$\mathcal{H}_2$ are denoted by $\Vert \cdot \Vert_1$ and $\Vert \cdot
\Vert_2$ respectively.

The algebraic tensor product
$\mathcal{H}_1 \otimes_{alg} \mathcal{H}_2$ of $\mathcal{H}_1$ and
$\mathcal{H}_2$ is
defined as usual as the set of all finite sums of the form $\sum_i p_i
\otimes q_i$ with $p_i \in \mathcal{H}_1$ and $q_i \in \mathcal{H}_2$.
We define a function $\mu$ on $\mathcal{H}_1 \otimes_{alg} \mathcal{H}_2$
by
\begin{equation} \label{mu}
\mu(t) := \inf \left\{ \sum_i \Vert p_i \Vert_1 \Vert q_i \Vert_2
\, \left\vert \, t = \sum_i p_i \otimes q_i \right. \right\}. \end{equation}
$\mu$ is a cross norm on $\mathcal{H}_1 \otimes_{alg} \mathcal{H}_2$
and the completion of $\mathcal{H}_1 \otimes_{alg} \mathcal{H}_2$
with respect to $\mu$ is a Banach
algebra which we denote by $\mathcal{H}_1 \otimes_{\mu} \mathcal{H}_2$
(for a proof, see, e.g., Proposition T.3.6 in \cite{WeggeOlsen93}).
The inner products on $\mathcal{H}_1$ and $\mathcal{H}_2$
induce an inner product on
$\mathcal{H}_1 \otimes_{alg} \mathcal{H}_2$ given by
\[ \langle a, b \rangle = \sum_{i,j} \langle p_i, t_j \rangle_1 \otimes
\langle q_i, s_j \rangle_2 \] if $a = \sum_i p_i \otimes q_i$ and $b =
\sum_j t_j \otimes s_j$. As \begin{eqnarray*}
\mu(\langle a, b \rangle) & = & \inf \left\{ \sum_{l} \Vert c_l
\Vert_1 \Vert d_l \Vert_2 \left\vert \langle a, b \rangle = \sum_l
c_l \otimes d_l \right. \right\} \\ & \leq & \mathrm{restr.}\inf
\sum_{i,j} \Vert \langle p_i, t_j \rangle_1 \Vert_1 \Vert
\langle q_i, s_j \rangle_2 \Vert_2  \\
& \leq & \mathrm{restr.}\inf
\sum_{i,j} \Vert p_i \Vert_1 \Vert q_i \Vert_2 \Vert t_j \Vert_1
\Vert s_j \Vert_2 \\ & = & \mathrm{restr.}\inf
\left( \sum_i \Vert p_i
\Vert_1 \Vert q_i \Vert_2 \right) \left( \sum_j \Vert t_j \Vert_1 \Vert
s_j \Vert_2 \right) \\
& = & \mu(a) \mu(b), \end{eqnarray*}
where the infimum in the first line runs over all possible
decompositions of $\langle a, b \rangle$ as sums over elementary
tensors, whereas the `restricted infima' in the following three lines
run over all
decompositions of $a$ and $b$ into sums of elementary tensors.
Consequently the inner product $\mu$ on
$\mathcal{H}_1 \otimes_{alg} \mathcal{H}_2$
is continuous and can be extended to the
completion $\mathcal{H}_1 \otimes_{\mu} \mathcal{H}_2$ of
$\mathcal{H}_1 \otimes_{alg} \mathcal{H}_2$.
We denote this extension also by $\mu$.
Therefore $\mathcal{H}_1 \otimes_{\mu} \mathcal{H}_2$
is a super Hilbert space when endowed with the norm $\mu$.

When both $\mathcal{H}_1$ and $\mathcal{H}_2$
satisfy the strong definiteness condition, both
$\mathcal{H}_1$ and $\mathcal{H}_2$ are
pre-Hilbert spaces with respect to the body of their inner
products. Therefore also the body $\mu_B$ of $\mu$ is a complex-valued
scalar product on $\mathcal{H}_1 \otimes_{alg} \mathcal{H}_2$
and, by virtue of the Cauchy-Schwarz
inequality, $\mu_B$ can be extended to
a complex-valued scalar product $\widetilde{\mu}_B$
on $\mathcal{H}_1 \otimes_{\mu} \mathcal{H}_2$.
$\widetilde{\mu}_B$ obviously coincides with the body of the
extension of $\mu$ to $\mathcal{H}_1 \otimes_{\mu} \mathcal{H}_2$.
Therefore we conclude that
$\mathcal{H}_1 \otimes_{\mu} \mathcal{H}_2$ is a super Hilbert space
satisfying the strong definiteness condition. \label{4.4}

In Section 6 we shall be interested in the case $\mathcal{H}_1 =
\Lambda_\infty^m$ and $\mathcal{H}_2 = \star \left[ \Lambda_\infty^m
\right]$. The norm $\mu_m$ arising from the mass norms on
$\Lambda^m_\infty$ and $\star \left[ \Lambda_\infty^m \right]$ via
Equation \ref{mu} is called the \textbf{mass norm} on
$\Lambda^m_\infty \otimes_{\mu_m} \star \left[
\Lambda_\infty^m \right]$.
It follows from our discussion above that $\Lambda^m_\infty
\otimes_{\mu_m} \star \left[ \Lambda_\infty^m \right]$ is a super
Hilbert space satisfying the strong definiteness condition.
We shall see in Section 6 that in the functional
Schr\"odinger representation
of spinor quantum field theory the super Hilbert space $\Lambda^m_\infty
\otimes_{\mu_m} \star \left[ \Lambda_\infty^m \right]$ arises
naturally as the quantum theoretical state space.
\end{ex}

\begin{prop}
Let $\mathcal{H}$ be a super Hilbert space and $T : \mathcal{H} \to
\mathcal{H}$ be an adjointable operator.
Then $T$ and $T^*$ are bounded with respect to the operator norm
\begin{equation}
\Vert T \Vert := \sup \{ \Vert Tx \Vert \, \vert \, \Vert x \Vert \leq 1
\}. \label{opnorm} \end{equation} If Equation \ref{knorm} holds,
then $\Vert T \Vert = \Vert T^* \Vert$. \label{4.8} \end{prop}
\emph{Proof}: Let $x_\lambda, x, y \in \mathcal{H}$, such that
$x_\lambda \to
x$ and $Tx_\lambda \to y$. The inner product of a super Hilbert
space is separately continuous in each variable. Thus
\[ 0 = \langle T^* e, x_\lambda \rangle - \langle T^* e, x_\lambda
 \rangle =
 \langle e, T x_\lambda \rangle - \langle T^* e, x_\lambda \rangle \to
 \langle e, y \rangle - \langle T^* e, x \rangle = \langle e, y - Tx
 \rangle, \] for all $e \in \mathcal{H}$. Putting $e = y - Tx$
 implies $y = Tx$. The boundedness of $T$ and $T^*$ follows now from
 the closed graph theorem. As $\Vert Tx \Vert^2 = \Vert \langle T^* T x, x
 \rangle \Vert \leq \Vert T^* T x \Vert \Vert x \Vert \leq \Vert T^* \Vert
 \Vert T \Vert \Vert x \Vert^2$, we find $\Vert T \Vert \leq \Vert
 T^* \Vert$. But then also $\Vert T^* \Vert \leq \Vert T^{**} \Vert =
 \Vert T \Vert$.   $\Box$ \\

 A similar argument proves
 \begin{prop}
Let $\mathcal{H}$ be a super Hilbert space satisfying the strong
definiteness condition
and $T : \mathcal{H} \to
\mathcal{H}$ be a weakly adjointable operator.
Then $T$ and $T^\dagger$ are bounded with respect to the operator
norm in Equation \ref{opnorm} and with respect to the norm
\begin{equation}
\Vert T \Vert_w := \sup \{ \vert \langle Tx, Tx \rangle_B
\vert^{1/2} \, \vert \, \Vert x \Vert \leq 1
\} \label{opnorm2} \end{equation}  and $\Vert T \Vert_w = \Vert T^\dagger
\Vert_w$. \end{prop}
\emph{Proof}: The boundedness of $T$ and $T^\dagger$ with respect
to the norm in Equation \ref{opnorm} follows as in the proof of
Proposition \ref{4.8}. The boundedness
with respect to $\Vert \cdot \Vert_w$ follows from $\Vert q \Vert_B
\leq \Vert q \Vert$ for all $q \in \Lambda$.  $\Box$
\begin{prop} Let $\mathcal{H}$ be a super Hilbert space.
When equipped with the operator norm
(\ref{opnorm}) $\mathfrak{B}(\mathcal{H})$ is an involutive
Banach algebra with continuous involution. \end{prop}
\emph{Proof}: It is easy to see that (\ref{opnorm}) defines a
norm on $\mathfrak{B}(\mathcal{H})$. The operator norm is clearly
submultiplicative. It remains to show that
$\mathfrak{B}(\mathcal{H})$ is norm complete.
If $(T_n)_{n \in \mathbb{N}}$ is a Cauchy sequence of
adjointable operators,  then $(T_nx)_{n \in
\mathbb{N}}$ and $(T^*_nx)_{n  \in \mathbb{N}}$ are Cauchy
sequences in $\mathcal{H}$ for every $x \in \mathcal{H}$.
We call the limits $Tx$ and $\overline{T}x$
respectively. Since $\langle y, Tx \rangle = \lim \langle y, T_n x \rangle
= \lim \langle T^*_ny, x \rangle = \langle \overline{T}y, x
\rangle,$ we see that $T$ is adjointable and $T^* = \overline{T}$.
This shows that $\mathfrak{B}(\mathcal{H})$ is
norm complete. From $\Vert T_n - T \Vert = \Vert T^*_n - T^* \Vert$
it is easy to see that the
involution is continuous.  $\Box$

\section{Physical observables}
\begin{de}
Let $E$ be a Hilbert $\Lambda$ module and $T \in
\mathfrak{B}_w(E)$. Then we say that a Gra\ss{}mann number $\lambda$
is a \textbf{spectral value} for $T$ when $T-\lambda I$ does not
have a two-sided inverse in $\mathfrak{B}_w(E)$. The set of spectral
values for $T$ is called the \textbf{spectrum} of $T$ and is denoted by
$\mathrm{sp}(T)$. The subset $\mathrm{sp}_\mathbb{C}(T) := \mathrm{sp}(T)
\cap \mathbb{C}$ is called the \textbf{complex spectrum} of $T$.
\end{de}

It is well-known that a Gra\ss{}mann number $q \in \Lambda_n, n$ finite,
has an inverse if
and only if its body $q_B$ is nonvanishing \cite{DeWitt92}.
Therefore the following proposition that the spectrum of a bounded
module map $T$ on a Hilbert $\Lambda_n$ module, $n$ finite,
is fully determined by the complex spectrum of $T$ is not surprising.
\begin{prop} \label{prop1}
Let $E$ be a
Hilbert $\Lambda_n$ module, $n$ finite, and $T \in
\mathfrak{B}_w(E)$ be a Hilbert module map.
Then $\lambda \in \mathrm{sp}(T)$ if and only
if $\lambda_B \in \mathrm{sp}_\mathbb{C}(T)$.
\end{prop}
\emph{Proof}: Let $\lambda \notin \mathrm{sp}(T)$. Then $T-\lambda
I$ has a two-sided inverse in $\mathfrak{B}_w(E)$, denoted by
$T_\lambda^{-1}$. Evidently $T_\lambda^{-1}$ is a module map.
Now let $s$ be a Gra\ss{}mann number with vanishing
body.
Then $T_{\lambda-s,L}^{-1} := \left( \sum_{n=0}^{\infty} (- T_\lambda^{-1}
s )^n \right) T_\lambda^{-1}$
is a left inverse for $T-(\lambda -s) I$ and $T_{\lambda-s, R}^{-1} :=
T_\lambda^{-1} \left( \sum_{n=0}^{\infty} (- s T_\lambda^{-1})^n \right)$
is a right
inverse for $T-(\lambda -s)I$. Both sums are actually finite. This
follows from the bodylessness of $s$ and from the fact that $T_\lambda^{-1}$
is decomposable
into an even and an odd part: $T_\lambda^{-1} =  T^{-1}_{\lambda,0} +
T_{\lambda,1}^{-1}$.
Therefore the left and right inverse exist for all $s \in
\Lambda_n$ with $s_B=0$.
As $T^{-1}_{\lambda-s,L} \left( T - (\lambda
-s) I \right) T_{\lambda-s,R}^{-1} = T^{-1}_{\lambda-s,R} =
T^{-1}_{\lambda -s,L}$ the left and right inverse coincide.
This proves that $\lambda \notin \mathrm{sp}(T)$ implies $\lambda -s \notin
\mathrm{sp}(T)$ for all $s \in \Lambda_n$ with $s_B =0$.  $\Box$

\begin{ex}
\label{ex2} Consider $\Lambda_n$ endowed with the inner product
(\ref{Eq2}). Let $\xi_1, \cdots, \xi_n$ denote the set of generators
of $\Lambda_n$. Consider the module map $\hat{\xi}_1 : \Lambda_n \to
\Lambda_n, \hat{\xi}_1 q := \xi_1 q$. Obviously $0$ is the only
complex spectral value of $\hat{\xi}_1$ and, as $\hat{\xi}_1 - s
I$ does not have an inverse for all bodyless $s \in \Lambda_n$,
all Gra\ss{}mann numbers with vanishing body are spectral values
for $\hat{\xi}_1$. The element $\xi_1 \cdots \xi_n \in \Lambda_n$
is an ``Eigenstate'' for $\hat{\xi}_1$ for any bodyless spectral
value: $\hat{\xi}_1 \xi_1 \cdots \xi_n =  s \xi_1 \cdots \xi_n =
0$, for all $s \in \Lambda_n$ with $s_B =0$.
\end{ex}

\begin{de}
Let $\mathcal{H}$ be a super Hilbert space. A \textbf{physical
observable} on $\mathcal{H}$ is a weakly self-adjoint operator
$\mathcal{O} : \mathcal{H} \to \mathcal{H}$. \end{de}
\begin{prop} Let $\mathcal{H}$ be a super Hilbert space and let
$H$ be the Hilbert space from Proposition \ref{3.3}. Then
there exists a * homomorphism $\varphi$ from
$\mathfrak{B}_w(\mathcal{H}) \cap \mathcal{L}(\mathcal{H})$
(equipped with the norm $\Vert \cdot \Vert_w$)
into the $C^*$-algebra $\mathfrak{B}(H)$
of bounded operators on $H$. \label{prop56} \end{prop}
\emph{Proof}: Let $\mathcal{N} := \{ x \in \mathcal{H} \vert \langle
x,x \rangle_B =0 \}$ and let $T \in \mathfrak{B}_w(\mathcal{H})$.
For $n \in \mathcal{N}$ we have by virtue of Lemma
\ref{lem1}: $\vert \langle T n, Tn \rangle_B
\vert^2 \leq \langle T^\dagger T n, T^\dagger
T n \rangle_B \langle n, n \rangle_B = 0$. Thus $T(\mathcal{N}) \subset
\mathcal{N}$. This shows that every $T \in
\mathfrak{B}_w(\mathcal{H})$ induces a bounded linear operator on
$\mathcal{H}/\mathcal{N}$ which we denote by $\phi(T)$ via
$\phi(T)(x + \mathcal{N}) := T(x) + \mathcal{N}$.
The operator $\phi(T)$
can be uniquely extended to a bounded linear operator
$\varphi(T)$ on $H$ (compare, e.g., Theorem 1.5.7 in \cite{KadisonR86}).
Obviously, the correspondence
$\varphi$ is linear, multiplicative and satisfies $\varphi(T^\dagger) =
\varphi(T)^*$ and $\varphi(I) = I_{\vert H}$, i.e., $\varphi$ is
a * homomorphism.  $\Box$

\begin{prop}
Let $\mathcal{H}$ be a super Hilbert space satisfying the strong
definiteness condition. Then the * homomorphism $\varphi$ from
Proposition \ref{prop56} is an isometric isomorphism from
$\mathfrak{B}_w(\mathcal{H})$ to the $C^*$-algebra
$\mathfrak{B}(H)$. Hence $\mathfrak{B}_w(\mathcal{H})$ is a
$C^*$-algebra with norm $\Vert T \Vert_{w} :=
\sup \{ \vert \langle Tx, Tx \rangle_B \vert^{1/2} \vert \Vert x \Vert
\leq 1 \}$.  \end{prop}
\emph{Proof}: This follows, e.g., from Theorem 1.5.7 in
\cite{KadisonR86}.  $\Box$

\begin{rema}
Proposition \ref{prop1} and Example \ref{ex2} show that in
general it is meaningless to
attribute physical relevance to the soul of a spectral value
of a physical observable. Instead it becomes clear that in general only the
elements of the complex spectrum of a physical observable may
admit an interpretation as possible physical values of the observable.
This is in accordance with and further substantiated by
the fact that Gra\ss{}mann numbers cannot be measured.
Above we have argued that the physical transition amplitudes on a
super Hilbert space are given by the body of the inner product.
Therefore it seems reasonable to define the \emph{physical
spectrum} of an operator $T$ on a super Hilbert space $\mathcal{H}$
as, loosely speaking, the subset of
$\mathrm{sp}_\mathbb{C}(T)$ corresponding to physical elements of
$\mathcal{H},$ i.e., \end{rema}
\begin{de}
The \textbf{physical spectrum} of a bounded
physical observable $\mathcal{O}$
on a super Hilbert space $\mathcal{H}$, denoted by
$\mathrm{sp}_{ph}(\mathcal{O})$, is the set of $\lambda
\in \mathbb{C}$ such that $\varphi(\mathcal{O}- \lambda I)$ has no two-sided
inverse in $\mathfrak{B}(H)$.
\end{de} \begin{prop} Let $\mathcal{O}$ be a bounded
physical observable on a super
Hilbert space $\mathcal{H}$.
Then $\mathrm{sp}_{ph}(\mathcal{O}) = \mathrm{sp}(\varphi(\mathcal{O}))
\subset \mathrm{sp}_\mathbb{C}(\mathcal{O})$. If
$\mathcal{H}$ satisfies the
strong definiteness condition, then $\mathrm{sp}_{ph}(\mathcal{O}) =
\mathrm{sp}_\mathbb{C}(\mathcal{O})$. \end{prop}
\emph{Proof}: Let $\lambda \notin \mathrm{sp}_\mathbb{C}(\mathcal{O})$,
$\lambda \in \mathbb{C}$.
Then $\mathcal{O} - \lambda I$ has a two-sided inverse in
$\mathfrak{B}_w(\mathcal{H})$ and as $\varphi$ is an
algebra-homomorphism, also $\varphi(\mathcal{O}) - \lambda I$ has a
two-sided inverse in $\mathfrak{B}(H)$. This implies $\lambda \notin
\mathrm{sp}(\varphi(\mathcal{O}))$.
The other direction follows analogously if $\varphi$
is an isomorphism.  $\Box$
\begin{co}
The physical spectrum of a bounded
physical observable on a super Hilbert space
is a compact subset of $\mathbb{R}$. \end{co}
If $\mathcal{H}$ is a super Hilbert space, $T \in
\mathcal{L}(\mathcal{H})$, then we say that an element $\lambda \in
\mathrm{sp}(T)$ is a \emph{right Eigenvalue} for $T$ if there is an $x \in
\mathcal{H}$ such that $T(x) = x \lambda$. $x$ is called an
\emph{Eigenvector} corresponding to $\lambda$. Notice that an Eigenvector
might correspond to more than one Eigenvalue, see Example
\ref{ex2}. \begin{co}
Let $\mathcal{O}$ be a bounded physical
observable $\mathcal{O}$
on a super Hilbert space. Let
$\lambda, \lambda' \in \mathrm{sp}_{ph}(\mathcal{O})$
be physical spectral values of $\mathcal{O}$
with $\lambda \neq \lambda'$
and let $x, x'$ be Eigenvectors
corresponding to $\lambda$ and $\lambda'$ respectively. Then $\langle
x, x' \rangle_B =0$. \end{co}
\begin{rema}
It is instructive to compare our definition of a physical
observable with the
definition given by DeWitt in \cite{DeWitt92}.
According to DeWitt's definition an element of a super Hilbert space is
called \emph{physical} if it has nonvanishing body.
A \emph{physical observable} is
then defined as a self-adjoint module map on the super Hilbert
space such that
\begin{itemize}
\item all Eigenvalues are even Gra\ss{}mann numbers;
\item for every Eigenvalue there is a physical Eigenstate;
\item the set of physical Eigenvectors that correspond to soulless
Eigenvalues contains a complete basis (for a definition of this notion see
\cite{DeWitt92}). \end{itemize}
Our argument above shows that DeWitt's additional assumptions are
unnecessary to assure real-valuedness of physical
spectral values and weak orthogonality
of Eigenvectors of physical observables. Moreover, Proposition \ref{prop1}
shows that for super Hilbert spaces over finitely generated
Gra\ss{}mann algebras there are no physical observables in
DeWitt's sense, as there will always be Eigenvalues which are not even.
For super Hilbert spaces over an infinitely
generated Gra\ss{}mann algebra it is also easy to see that in
general DeWitt's conditions cannot be satisfied. For let $T$ be a
physical observable in DeWitt's sense with Eigenvalue $\lambda$, and let
$x$ denote a physical Eigenstate for $\lambda$. Then $\lambda +
\xi_1$ is an Eigenvalue of $T$ with Eigenstate $x \xi_1 \neq 0$.
The Eigenvalue  $\lambda + \xi_1$ is neither an even Gra\ss{}mann
number nor it is guaranteed that there is a physical Eigenstate
for this Eigenvalue. This is a contradiction. Therefore we
conclude that in general there are no physical observables in
DeWitt's sense on a super Hilbert space.
\end{rema}

\section{The Schr\"odinger representation of spinor quantum field
theory}
In the Schr\"odinger representation of quantum field theory the
commutation relations of the field operators are realized by
representing the field operators by
functionals and representing the conjugate momenta by
functional derivatives \cite{Hatfield92}. This formulation of quantum
field theory is equivalent to the standard operator formulation
and to the functional-integral representation of quantum field
theory. In the Schr\"odinger representation of spinor quantum
field theory super Hilbert spaces naturally arise as
the quantum mechanical state space. The material presented in this
section is taken from Hatfield \cite{Hatfield92}.

The Hamiltonian for free spinor field theory is given by
\begin{eqnarray*}
H & = & \int d^3 x \Psi^\dagger(x)(- i \vec{\alpha} \cdot \nabla +
\beta m) \Psi(x), \\
& = & \int d^3 x \bar{\Psi}(x)(- i \gamma^k \partial_k +m) \Psi(x)
\end{eqnarray*}
where the matrices $\alpha_i$ and $\beta$ satisfy
$\beta^2 = 1, \{\alpha_i, \alpha_j \} =  2 \delta_{ij}, \{ \alpha_i,
\beta \} =0$ and the $\gamma$ are related by $\gamma^0 = \beta$
and $\alpha_i = \gamma^0 \gamma_i$.

The canonical anticommutation relations of the fields are given by
\begin{eqnarray*}
\{\Psi_\alpha(\vec{x},t), \Psi^\dagger_\beta(\vec{y},t) \} & = &
\delta_{\alpha \beta} \delta^3(\vec{x} - \vec{y}) \\
\{ \Psi_\alpha(\vec{x}, t), \Psi_\beta(\vec{y},t) \} & = & \{
\Psi^\dagger_\alpha(\vec{x},t), \Psi^\dagger_\beta(\vec{y}, t) \}
=0. \end{eqnarray*}

The time evolution is given by the functional Schr\"odinger
equation \[ i \frac{\partial}{\partial t} \vert \Psi \rangle =
H \vert \Psi \rangle. \]

In the ``coordinate'' Schr\"odinger representation the state space
at time $t$
is spanned by the Eigenfunctions $\vert \Psi \rangle$ of the field
operator $\Psi(x)$. The corresponding Eigenvalues $\psi(x)$ are
spinors of Gra\ss{}mann number-valued
functions. The conjugate momentum operator
$\Psi^\dagger(\vec{x})$ is represented as a functional derivative
\[ \Psi^\dagger_\beta(\vec{x}) = \frac{\delta}{\delta
\psi_\beta(\vec{x})}. \]
It is well-known that the Eigenfunctions and the
functional derivative can be rewritten as a plane wave expansion
in terms the creation and annihilation operators
\begin{eqnarray*}
\psi_\alpha(\vec{x}) & = & \sum_{i=1}^2 \int \frac{d^3 p}{\sqrt{(2
\pi)^3}} \sqrt{\frac{m}{E}} \left[ b_i(\vec{p}) u^i_\alpha(\vec{p}) e^{-i
\vec{p} \vec{x}} + d^\dagger_i(\vec{p})
v^i_\alpha(\vec{p}) e^{i
\vec{p} \vec{x}} \right] \\
\frac{\delta}{\delta \psi_\alpha(\vec{x})} & = &
\sum_{i=1}^2 \int \frac{d^3 p}{\sqrt{(2
\pi)^3}} \sqrt{\frac{m}{E}} \left[ \frac{\delta}{\delta
b_i(\vec{p})} (u^i_\alpha)^\dagger(\vec{p}) e^{i
\vec{p} \vec{x}} + \frac{\delta}{\delta d^\dagger_i(\vec{p})}
(v^i_\alpha)^\dagger(\vec{p}) e^{-i
\vec{p} \vec{x}} \right],
\end{eqnarray*}
where
\begin{equation}
u^i(\vec{p}) = \frac{\slash \!\!\! p +m}{\sqrt{2m (m +E)}} \left(
\begin{array}{c}
\delta_{i1} \\ \delta_{i2} \\ 0 \\ 0
\end{array} \right), \; v^i(\vec{p})  =
\frac{- \slash \!\!\! p +m}{\sqrt{2m (m +E)}} \left(
\begin{array}{c}
0 \\ 0 \\ \delta_{i1} \\ \delta_{i2}
\end{array} \right).
\end{equation}
The operators $b(\vec{p}), \frac{\delta}{\delta b(\vec{p})},
d^\dagger(\vec{p})$ and
$\frac{\delta}{\delta d^\dagger(\vec{p})}$ act on the state space
and obviously satisfy the equal
time anticommutation relations
\begin{eqnarray*}
\left\{b_i(\vec{p}), \frac{\delta}{\delta b_j(\vec{k})} \right\}
& = & \delta^3(\vec{p} - \vec{k}) \delta_{ij} \\
\left\{ \frac{\delta}{\delta d^\dagger_i(\vec{p})},
d^\dagger_j(\vec{k}) \right\} & = &
\delta^3(\vec{p} - \vec{k}) \delta_{ij}
\end{eqnarray*} with all other anticommutators vanishing.
Accordingly, the $\frac{\delta}{\delta b_i}$ and the
$\frac{\delta}{\delta d^\dagger_i}$ are interpreted as
creation operator of field quanta with positive or negative energy
respectively whereas the $b_i$ and the $d^\dagger_i$ are
interpreted as the corresponding annihilation operators.

At this stage
the important observation for our
purposes is that the state space can be naturally identified as
the tensor product super Hilbert space $\widetilde{\Lambda}^m_\infty \equiv
\Lambda^m_{\infty,d} \otimes_\mu \star \left[\Lambda_{\infty, b}^m \right]$
where the uncountable set of generators of the first factor
($\Lambda_{\infty,d}^m$) is identified with
$\{ d_i^\dagger(\vec{p}) \}$ and the set of generators
of $\Lambda^m_{\infty, b}$ is identified with $\{ b_i(\vec{p}) \}$,
see Example \ref{4.4}. As all generators of $\Lambda^m_{\infty,d}$
commute with all generators of $\Lambda^m_{\infty,b}$, we omit the
tensor symbol in our notation.
A factor $b_i(\vec{p})$ (or $d_i^\dagger(\vec{p})$) in an
element $x \in \widetilde{\Lambda}^m_\infty$ corresponds to the
situation that the field quantum annihilated by $b_i(\vec{p})$ (or
$d_i^\dagger(\vec{p})$) is \emph{absent}. For example, the vacuum
state, where all negative energy states are filled and all
positive energy states are empty, is represented by
\[ \vert 0 \rangle = \prod_{i=1}^2 \prod_{\vec{p}} b_i(\vec{p}) =
\xi_{\infty,b} \]
and the state with one positron of momentum $\vec{p}_p$ with spin
up is given by \[ \vert \vec{p}_p, \uparrow \rangle  =
d_2^\dagger(\vec{p}_p) \prod_{i=1}^2 \prod_{\vec{p}} b_i(\vec{p}). \]
Physical transition amplitudes are computed by performing a functional
integration over all Gra\ss{}mann degrees of freedom, see
\cite{Hatfield92}. Mathematically this is equivalent to taking the
body of the inner product (\ref{Eq2}) as the physical transition
amplitude, compare Proposition \ref{3.3}. It is now obvious that
-- when identifying the Hilbert space $H$ in Proposition
\ref{3.3} with the state space in the operator (Heisenberg)
representation of spinor quantum field theory -- these physical
transition amplitudes in the Schr\"odinger representation coincide
with the physical transition amplitudes in the operator
representation.

\section{Previous definitions revisited}
\subsection{DeWitt super Hilbert spaces}
Super Hilbert spaces were first considered by DeWitt in his book
\cite{DeWitt92}. The basic features of DeWitt's definition may
be summarized as follows: DeWitt defines a super Hilbert space $\mathcal{H}$
basically as a $\mathbb{Z}_2$-graded $\Lambda_n$ module,
where $n$ is possibly infinite, with a $\Lambda_n$-valued
inner product $\langle \cdot, \cdot \rangle : \mathcal{H} \times
\mathcal{H} \to \Lambda_n$ subject to the following conditions
\begin{enumerate}
\item $\langle x, y_1 + y_2 \rangle = \langle x, y_1 \rangle + \langle x,
y_2 \rangle$, for $x,y_1,y_2 \in \mathcal{H}$;
\item $\langle x, \alpha y \rangle = \alpha \langle x, y \rangle =
\langle \alpha^* x, y \rangle$,
for $x,y \in \mathcal{H}, \alpha \in \mathbb{C}$;
\item $\langle x, y q \rangle = \langle x, y \rangle q$
for all $x,y \in \mathcal{H}, q \in \Lambda_n$.
\item $\langle x, y \rangle = \langle y, x \rangle^*$, for $x,y \in
\mathcal{H}$;
\item $\langle x, x \rangle_B \geq 0$ for $x \in \mathcal{H}$;
$x \in \mathcal{H}$ has nonvanishing body if and only if $\langle x, x
\rangle_B > 0$;
\item $\langle x_s, y_r \rangle q_t = (-1)^{t(s+r)} q_t \langle x_s, y_r
\rangle$
for all pure $x_s \in \mathcal{H}_s, y_r \in \mathcal{H}_r$ and
$q \in \Lambda_n$, $\deg(q_t)=t$.
\end{enumerate}
DeWitt moreover requires that the body of $\mathcal{H}$ is an
ordinary complex Hilbert space. The central difference between our
definition and DeWitt's is the requirement of sesqui-$\Lambda$-linearity of
the inner product. It is easy to see that sesqui-$\Lambda$-linearity
implies $\langle x_B, y_B
\rangle = \langle x, y \rangle_B$ for all $x, y \in \mathcal{H}$.
DeWitt concentrated on algebraic properties of super Hilbert
spaces and did not consider the topological or metric
structure of $\Lambda_n$ or $\mathcal{H}$.

Nagamachi and Kobayashi formalized and refined DeWitt's definition
by taking also into account the topological and norm structure on super Hilbert
spaces \cite{NagamachiK92}. It becomes clear from their work that
the requirement of sesqui-$\Lambda$-linearity in some sense
trivializes the theory, as it is possible to show along their lines
that, when $\Lambda_n$ is equipped with the Rogers norm, every
super Hilbert space is of the form $\mathcal{H} = H \otimes
\Lambda_n$ where $H$ is an ordinary complex Hilbert space.
\subsection{El Gradechi and Nieto's super Hilbert space}
El Gradechi and Nieto studied in \cite{ElGradechiN96} a super
extension of the Kirillov-Kostant-Souriau geometric quantization method.
They defined a super Hilbert space to be
a direct sum $\mathcal{H} = H_0 \oplus H_1$ of two complex Hilbert spaces
$(H_1, \langle \cdot, \cdot \rangle_0)$ and $(H_2, \langle \cdot, \cdot
\rangle_1)$ equipped with the super Hermitean form $\langle \! \langle
\cdot, \cdot \rangle \! \rangle = \langle \cdot, \cdot \rangle_0 +
i \langle \cdot, \cdot \rangle_1$.

At first sight this definition looks rather different from the
approach given in this paper. However, El Gradechi's and Nieto's
definition is
actually abstracted from the concrete example arising in their study
of super unitary irreducible representations of $\mathrm{OSp}(2/2)$
in super Hilbert spaces of $L^2$ superholomorphic sections of
prequantum bundles of the Kostant type. It is beyond the scope of
this paper to review the construction in \cite{ElGradechiN96} in
detail. For our purposes it is enough to know that the super
Hilbert space $\mathcal{L}_p^{(1 \vert 2)}$ constructed in
\cite{ElGradechiN96} is a
$\mathbb{Z}_2$-graded $\Lambda_4$ module and that its elements
are sections of the form $\psi(z, \overline{z}, \theta, \overline{\theta},
\chi, \overline{\chi})$, where $z \in \mathbb{C}$ denotes a
complex variable with $ \vert z \vert \leq 1$
and $\theta, \overline{\theta}, \chi, \overline{\chi}$
denote a complexified set of generators of $\Lambda_4$. Notice that
the definition for complex conjugation $\overline{ }$ in $\Lambda_4$
adopted in \cite{ElGradechiN96} differs from our definition:
$\overline{pq} = \overline{p} \, \overline{q}$ for all $p,q \in \Lambda_4$.
The even super Hermitean form on $\mathcal{L}_p^{(1 \vert 2)}$ is
defined for $\psi = \psi_0 + \psi_1$ and
$\psi' = \psi_0' + \psi_1'$ by (see Equation 5.8 and 5.10 in
\cite{ElGradechiN96})
\[ \langle \! \langle \psi', \psi \rangle \! \rangle
\equiv \int (\psi', \psi)
h d z d \overline{z} d \theta d \overline{\theta}
d \chi d \overline{\chi}, \]
where $h$ is an integrating constant factor for the super Liouville measure
constructed in \cite{ElGradechiN96} and
where \[ (\psi', \psi) = \overline{\psi'} \psi =
\overline{\psi_0'} \psi_0 + \overline{\psi_1'}
\psi_1 + \overline{\psi_0'} \psi_1 + \overline{\psi_1'} \psi_0. \]
It is now crucial to realize that it is possible to construct an
$\Lambda_4$-valued inner product $\langle \cdot, \cdot \rangle$
on $\mathcal{L}_p^{(1 \vert 2)}$ in the sense of
Definition \ref{def2} by replacing $(\psi', \psi)$ by
\[ (\psi', \psi)^\sim = {\psi'}^* \psi
\] (where $^*$ denotes the complex conjugation operation introduced
in Section 2) and by setting \begin{equation} \label{IP}
\langle \psi', \psi \rangle \equiv \int (\psi', \psi
)^\sim h dz d \overline{z}. \end{equation}

This is an inner product of the form
Equation \ref{kinner} and therefore
the physical super Hilbert space can be identified with
the super Hilbert space discussed in Example \ref{exa}.
Notice that what is called a \emph{super unitary operator} with respect to
the super Hermitean product in
\cite{ElGradechiN96} would be called a \emph{weakly unitary
operator} with respect to the inner product (\ref{IP}) in our terminology.
El Gradechi and Nieto correctly
note that the examples for super Hilbert spaces considered by them
are not covered by the definitions of DeWitt, and of Nagamachi and
Kobayashi.
\subsection{Other definitions}
There are some other notions of super Hilbert space in the
literature which we briefly mention here. Khrennikov \cite{Khrennikov91}
defines a super Hilbert space to be a Banach (commutative) $\Lambda$
module which is isomorphic to the space $\ell_2(\Lambda)$
of square-summable sequences in $\Lambda$
with the inner product $\langle x, y \rangle :=
\sum x_n y_n^*$ and norm $\Vert x \Vert^2 := \langle x, x
\rangle$.

Schmitt proposed a different notion of super Hilbert
space in \cite{Schmitt90}. According to his definition a super
Hilbert space is just a complex $\mathbb{Z}_2$-graded ordinary Hilbert
space (consequently the inner product is always complex-valued).
It is beyond the scope of the present work to discuss this
definition in detail and the reader is referred to
\cite{Schmitt90} and references therein.
We only remark that
Schmitt's definition of super Hilbert space is general enough
to cover even the functional Schr\"odinger representation of
spinor quantum field theory and contact to our approach can be
made for instance
by identifying his ``super Hilbert space'' with the Hilbert space
of Proposition \ref{3.3}.
\subsection{Discussion}
The definitions of the notion of super Hilbert space put forward
by DeWitt \cite{DeWitt92}, Nagamachi and Kobayashi \cite{NagamachiK92} and
Khrennikov \cite{Khrennikov91}
are all special cases of our more general definition.
From a
mathematical point of view they are viable generalizations of ordinary
Hilbert space theory, and indeed, unlike the definition of super Hilbert
space introduced in this paper, the definitions by Nagamachi,
Kobayashi and Khrennikov are actually designed such that analogies
or extensions
of most basic structural results of ordinary Hilbert space theory
remain valid in the super case.
However, these definitions suffer from the problem that they are too narrow
to cover the physically important example of
the state space arising in the functional Schr\"odinger
representation of spinor quantum field theory discussed in Section
6. As repeatedly stated above,
in physical applications of super Hilbert spaces the
physical transition amplitudes have to be identified with the body
of the inner product (this is in accordance with DeWitt's remark that
``real physics is restricted to the ordinary Hilbert space that sits inside
the super Hilbert space'').
However, in the theories by DeWitt, Nagamachi,
Kobayashi and Khrennikov the inner product respects the body
operation $\langle x_B, y_B \rangle =
\langle x, y \rangle_B$. In the functional Schr\"odinger representation
of spinor quantum field theory the physical state space is identified
with the tensor product of two isomorphic copies of the
underlying Gra\ss{}mann algebra: the presence of, e.g., a Gra\ss{}mann
number $d^\dagger_i(\vec{p})$ indicates the presence of the
corresponding positron field quantum. Therefore the body of a DeWitt-type
inner product gives essentially only the
contribution of the vacuum-to-vacuum transition amplitude to
the full physical transition amplitude. To obtain the
full physical transition amplitude we need to introduce the
more general inner product on the state space, given by Equation
\ref{Eq2}, which is not sesqui-$\Lambda$-linear and for which in
general $\langle x_B, y_B \rangle \neq \langle x, y \rangle_B$.
The super Hilbert spaces arising in the work of El Gradechi and
Nieto \cite{ElGradechiN96} and of Samsonov \cite{Samsonov97}
provide further examples for super Hilbert spaces which are not
super Hilbert spaces in the sense of DeWitt, Nagamachi and
Kobayashi, or Khrennikov but which are super Hilbert spaces in the sense
of Definition \ref{SH}. Therefore we
conclude that the introduction of the more general notion of super Hilbert
space put forward in this paper is physically justified.

\subsection*{Acknowledgements}
This research was performed while the author was
a Marie Curie Research Fellow in the Theoretical Physics Group
at Imperial College, London, UK,
under the Training and Mobility of Researchers (TMR)
programme of the European Commission.

\end{document}